\documentclass[aps,pra,twocolumn,english,showpacs,superscriptaddress]{revtex4-1}
\usepackage{graphicx}
\usepackage{amssymb}
\usepackage{amsmath, amstext, amssymb, amsfonts, amsxtra, mathtools}  %, stmaryrd}
\usepackage{braket}
\usepackage{textcomp}
\usepackage{relsize}
\usepackage[usenames,dvipsnames]{color}
\usepackage{units}
\usepackage[normalem]{ulem}

\newcommand{\aver}[1]{\langle #1 \rangle}

\newcommand{\bonn}{HISKP, University of Bonn, Nussallee 14-16, 53115 Bonn, Germany}
\newcommand{\beheshti}{Department of Physics, Shahid Beheshti University, G.C., Evin, Tehran 19839, Iran} 

\begin{document}

%==============================================================================================
\begin{abstract}
We propose two experimental setups for fermionic atoms in a high-finesse optical resonator in which either a superconducting state with s-wave symmetry of the pairs or a 4$k_F$ charge density wave can self-organize. In order to stabilize the s-wave pairing, a two component attractively interacting fermionic gas is confined to a one dimensional chain structure by an optical lattice. The tunneling of the atoms along the chains is suppressed initially by an energy offset between neighboring sites. A Raman transition using the cavity mode and a transversal pump laser then reintroduces a cavity-assisted tunneling. The feedback mechanism between the cavity field and the atoms leads to a spontaneous occupation of the cavity field and of a state of the fermionic atoms which is dominated by s-wave pairing correlations. Extending the setup to a quasi-one-dimensional ladder structure where the tunneling of atoms along the rungs of the ladder is cavity-assisted, the repulsively interacting fermionic atoms self-organize into a $4k_F$ charge density wave. We use adiabatic elimination of the cavity field combined with state-of-the-art density matrix renormalization group methods in finite systems in order to identify the steady state phases of the system. 
\end{abstract}

\title{Cavity-induced superconducting and 4$k_F$ charge-density-wave states} 

\author{Ameneh Sheikhan}
\affiliation{\beheshti}
\affiliation{\bonn}
\author{Corinna Kollath}
\affiliation{\bonn}

\maketitle
%==============================================================================================
%introduction
\section{Introduction}

The coupling of light to matter has been the subject of exciting investigation. In particular, the coupling of ultracold atoms to the light field of an optical resonator has lead to many interesting phenomena \cite{RitschEsslinger2013}. This includes the realization of the Jaynes Cummings model by the coupling of a single two-level atom to the radiation field of an optical high finesse cavity \cite{Agarwel, HarocheRaimond} and  the recent open-system realization of the Dicke model by the loading of a Bose-Einstein condensate into an optical cavity \cite{BaumannEsslinger2010,KesslerHemmerich2014}. The Dicke model exhibits a quantum phase transition due to a collective coupling of the atoms to the light field; the condensed atoms can be either in a phase characterized by a homogeneous density and a vanishing cavity field or self-organize into a supersolid phase with checkerboard density patterns and occupied cavity field. This phase transition which was predicted theoretically \cite{DomokosRitsch2002, DimerCarmichael2007, NagyDomokos2008, PiazzaZwerger2013} is driven by the cavity induced long-range interactions between atoms. In the open-system realization the self-organized phase is the so-called attractor state of the dissipative system due to the photon loss from the mirrors of the cavity.

The additional application of underlying static lattices has been achieved in order to investigate the direct competition between short-range interactions of the atoms and the cavity induced long-range interactions \cite{ElliotMekhov2016, BakhtiariThorwart2015,MaschlerRitsch2005, MaschlerRitsch2008, LarsonLewenstein2008, NiedenzuRitsch2010, SilverSimons2010, VidalMorigi2010, LiHofstetter2013}. The interplay between these interactions leads to the emergence of complex phases e.g.~a self-organized Mott-insulator or a self-organized charge density wave state \cite{LandigEsslinger2016,KlinderHemmerich2015}.
A variety of different self-organized phases has been proposed theoretically \cite{RitschEsslinger2013}. For example, the self-organization of bosonic atoms into different geometries \cite{SafaeiGremaud2015,LeonardDonner2017, LeonardDonner2017}, of fermionic atoms into super-radiant phases \cite{LarsonLewenstein2008b,MuellerSachdev2012,PiazzaStrack2014,KeelingSimons2014,ChenZhai2014,ChenYu2015} and in multimode cavities  \cite{GopalakrishnanGoldbart2009, NimmrichterArndt2010, StrackSachdev2011, GopalakrishnanGoldbart2011, HabibianMorigi2013, JanotRosenow2013, BuchholdDiehl2013} have been studied. 

Moreover, different theoretical proposals have been designed in order to realize a  spin-orbit coupling \cite{DengYi2014,DongPu2014,PanGuo2015,PadhiGosh2014,MivehvarFeder2014,MivehvarFeder2015} and artificial gauge fields \cite{HalatiKollath2017, KollathBrennecke2016, SheikhanKollath2016, SheikhanKollath2016R, WolffKollath2016, ZhengCooper2016, BallantineKeeling2017} mediated by the cavity field. In these systems the self-organized phases can be topologically non-trivial and carry chiral current. 

In this work interacting fermions are coupled to a cavity mode exploiting Raman processes involving a
cavity mode to induce tunneling between two sites of a preexisting lattice. We show the self-organization of fermions into an s-wave superconducting state in case of one dimensional chains with attractive interaction and into a charge density wave with $4k_F$-oscillations in the case of a ladder geometry with repulsive interaction.

In the following we first describe the setup for the interacting fermions in one-dimension coupled to a cavity mode and introduce the corresponding theoretical model in Sec.~\ref{sec:setup1d}. Particularly, by adiabatically eliminating the cavity mode we derive an effective Hamiltonian for the fermionic atoms together with a self-consistency condition. We also give the stability condition for the non-trivial self-organized solutions of the effective Hamiltonian in which the cavity is occupied. In Sec.~\ref{sec:swave} we study the system solving the self-consistent equation numerically by DMRG methods. Also the stability of the s-wave superconducting phase induced by the cavity is investigated. In Sec.~\ref{sec:setupq1d} we introduce the fermionic system in the ladder geometry coupled to the cavity. In Sec.~\ref{sec:cdw} we discuss the properties and stability of self-organized phases of this model.

%Description of the setup
\section{Setup and theoretical model for the one-dimensional chains}\label{sec:setup1d}

\begin{figure*}
  \includegraphics[width=0.7\linewidth]{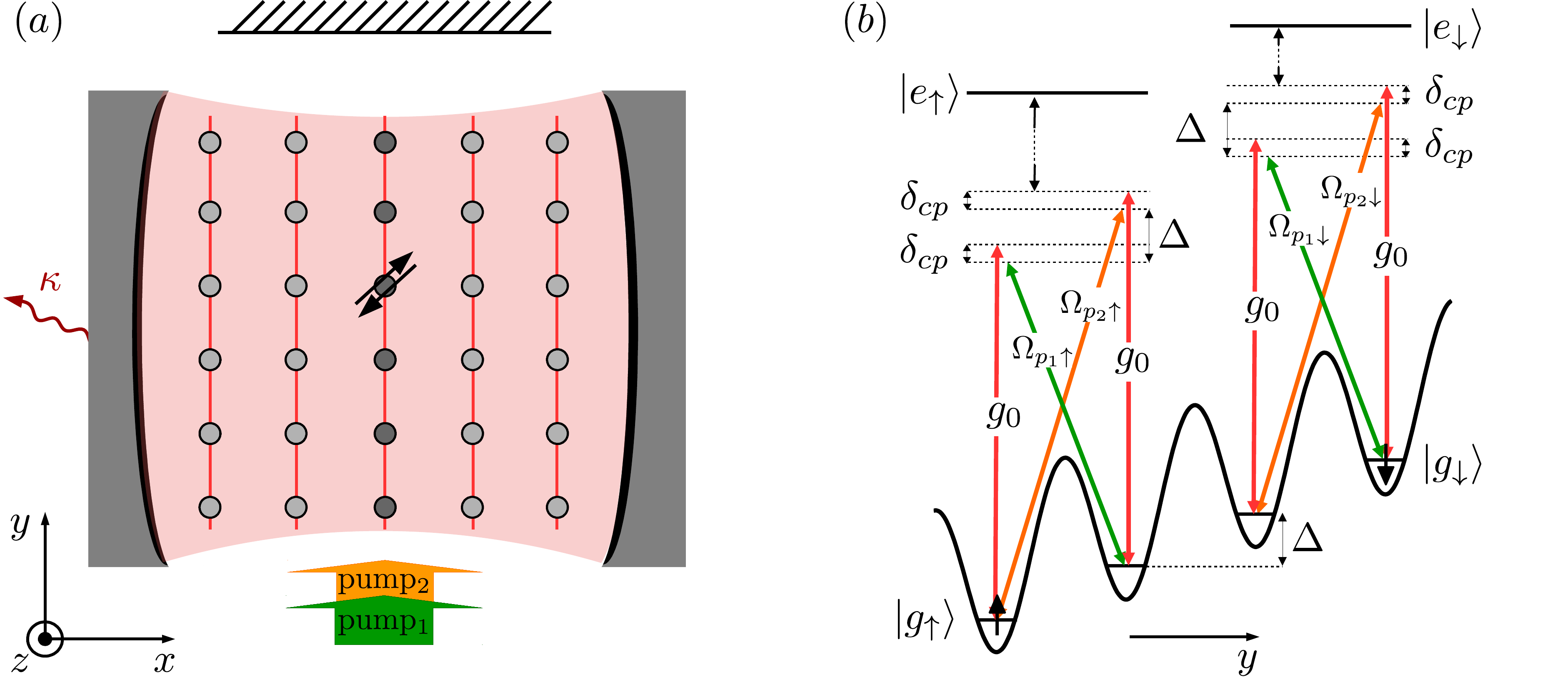}
\caption{\label{fig:setup}(color online) (a) Two different  internal states of fermionic atoms represented by spin $\uparrow$ and $\downarrow$ are loaded into decoupled one-dimensional structures formed by optical lattices and coupled to the dynamical field of an optical cavity. Tunneling along the chain ($y$-direction) is strongly suppressed by a potential offset $\Delta$ between neighboring lattice sites. (b) Cavity-assisted Raman processes induced by two running-wave pump beams, restore tunneling along the $y$-direction for each spin state of the atoms. The ground states are $|g_\uparrow\rangle$ and $|g_\downarrow\rangle$ and the intermediate states of the Raman process are $|e_\uparrow \rangle$ and $|e_\downarrow \rangle$, respectively. The two pump laser beams are coupled differently to spin $\sigma=\uparrow,\downarrow$ with Rabi frequencies $\Omega_{p_1 \sigma},\Omega_{p_2 \sigma}$ and $g_0$ is the vacuum-Rabi frequency of the cavity. The cavity-pump detuning is denoted by $\delta_{cp}$.} 
\end{figure*}

We consider an ultracold spin-balanced interacting fermionic gas inside an optical cavity, where the cavity is oriented along $x$-direction (see Fig.~\ref{fig:setup}(a)). The fermions are subjected to an anisotropic, three-dimensional optical lattice which confines them into an array of decoupled chains along the $y$-direction. %This is achieved using a strong optical lattice along $x$ and $z$-directions. 
Tunneling in the $y$-direction is suppressed by an energy offset $\Delta$ between neighboring sites which, for example, can be created by an AC Stark shift gradient or a superlattice.
For both spin states, tunneling is restored by two resonant two-photon Raman transitions similar to the setups discussed in Ref.~\cite {MiyakeKetterle2013} for the gradient and in Ref.~\cite{AidelsburgerBloch2013} for the superlattice. We describe here the situation for the gradients in detail which can then be adapted to the superlattice setup. 
The Raman transitions use two standing-wave pump laser beams with frequencies $\omega_{p_1},\omega_{p_2}$ along $y$-direction and one mode of the cavity with resonance frequency $\tilde{\omega}_c \approx \omega_{p_1} + \Delta/\hbar \approx \omega_{p_2} - \Delta/\hbar$ (see Fig.~\ref{fig:setup} b). We assume all other cavity modes to be far-detuned. 
The two pump laser beams are coupled differently to two spin states with Rabi frequencies $\Omega_{p_1 \sigma},\Omega_{p_2 \sigma}$ where $\sigma = \uparrow,\downarrow$. The intermediate states of the Raman process for each spin are $|e_\uparrow \rangle$ and $|e_\downarrow \rangle$ with internal atomic transition $\omega_{e\uparrow}$ and $\omega_{e\downarrow}$. The detuning of the pump and cavity modes remain large i.e. $\omega_{e\uparrow},\omega_{e\downarrow}\gg \tilde{\omega}_c,\omega_{p_1},\omega_{p_2}$ so that excitation by single-photon absorption are negligible in comparison to the coherent transfer from one site to the neighboring site. We can eliminate the excited state adiabatically and describe the system in terms of the atomic spin up and spin down states and the cavity mode. Considering the periodic structure of the underlying lattice we can express the Hamiltonian in the tight-binding approximation. Using the rotating frame with frequency $\omega_p = \frac{\omega_{p_1}+\omega_{p_2}}{2} $, we obtain
\begin{eqnarray}\label{eq:Ham_i}
&&H = H_c + H_{ac} + H_{\text{int}}, \nonumber\\
&&H_c=\hbar \delta_{cp} a^\dagger a, \nonumber\\
&&H_{ac}= - \hbar\tilde{\Omega} (a^\dagger+a) \sum_{j,\sigma} \left (c^\dagger_{j,\sigma} c_{j+1,\sigma}+\mathrm{H.c.} \right )\nonumber\\
&&H_{\text{int}}= U \sum_j n_{j,\uparrow} n_{j,\downarrow}.
\end{eqnarray}
Here, $c_{j,\sigma}$ $(c_{j,\sigma}^{\dag})$ annihilates (creates) a fermionic atom with spin $\sigma=\uparrow,\downarrow$ on site $j$ of the chain. The operator $n_{j,\sigma} = c_{j,\sigma}^{\dag} c_{j,\sigma} $ is the density operator. The fermions interact with the on-site interaction strength $U$. The cavity field operator $a$ ($a^\dag$) annihilates (creates) a cavity photon in the considered cavity mode and $H_c$ represents the dynamics of the cavity field in the rotating frame. The average cavity-pump detuning is denoted by $\delta_{cp}=\left(\tilde{\omega}_c-\omega_p \right)$.

%-----------------------------------------------------------------------------------------------
% Fig. 01
%-----------------------------------------------------------------------------------------------
%----------------------------------------------------------------------------------------------- 
The cavity-assisted tunneling along the chain is described by the term $H_{ac}$ with the effective Rabi frequency $\hbar\tilde{\Omega}=\frac{\hbar\Omega_{p_1\uparrow} g_0}{\omega_{e\uparrow}-\omega_{p_1}}\phi_{\parallel}\phi_{\perp}$. $\Omega_{p_1\uparrow}$ denotes the Rabi frequency of the first pump beam for the spin up state and $g_0$ the vacuum-Rabi frequency of the cavity. The overlap integrals $\phi_\parallel$ and $\phi_{\perp}$ are effective parameters depending on the Wannier states and can be tuned via the geometry of the optical lattice and the cavity mode \cite{KollathBrennecke2016}. In order to balance the strength of the tunneling in two directions along the chain for both spin up and down states, the Rabi frequency of the pump beam $i= 1,2$ for spin $\sigma=\uparrow,\downarrow$ is chosen as $\Omega_{p_i\sigma}=\frac{\Omega_{p_1\uparrow} (\omega_{e\sigma}-\omega_{p_i})}{(\omega_{e\uparrow}-\omega_{p_1})}$.

Considering the dissipative nature of the imperfect optical resonator, the dynamics of the system follows the Lindblad master equation, 
\begin{align}\label{eq:Lindblad}
&\dot \rho(t) = -\frac{i}{\hbar} \left[ H, \rho(t) \right] + \mathcal{D}(\rho(t)) \nonumber\\
&\mathcal{D}(\rho) =  \kappa \left( 2a \rho a^\dagger - \rho a^\dagger a - a^\dagger a \rho  \right),
\end{align}
where $\kappa$ is the bare cavity loss rate. The resulting equation of motion for the cavity mode is given by
\begin{align}\label{eq:a_time}
&i \partial_t a=-\tilde{\Omega} K +(\delta_{cp}- i \kappa ) a.
\end{align}
Here, $K=\sum_{j,\sigma} \left(c^\dagger_{j,\sigma} c_{j+1,\sigma}+\mathrm{H.c.} \right )$ is defined as the tunneling operator. 
Further, we eliminate the cavity field dynamics adiabatically from the equations of motion by using its steady state solution $\alpha$ of the equation~$\partial_t \langle a\rangle=0$ which is given by
\begin{equation}\label{eq:a_stat}
\alpha=\langle a \rangle = \frac{\tilde{\Omega}}{\delta_{cp}- i \kappa }\langle K \rangle.
\end{equation}
Replacing the cavity operator by its stationary state value in the equation of motions for the fermionic operators leads us to an effective Hamiltonian for the fermions which is the well known Hubbard model:
\begin{eqnarray}\label{eq:Ham_F}
&&H_{F} = - J K + H_{\text{int}}.
\end{eqnarray}
 The feedback of the cavity field on the dynamics of the atoms appears through the self-consistent determination of the tunneling amplitude $J=A\langle K \rangle$ with $A= \frac{2\hbar\tilde{\Omega}^2\delta_{cp}}{\delta_{cp}^2 +\kappa^2}$. We will call $A$ loosely the pump strength, since this is typically one of the easiest experimental nobes in order to tune the value of $A$. The stationary states of the system are determined by the solution of the effective Hamiltonian (Eq.~\ref{eq:Ham_F}) together with the self-consistency equation. In order to determine the steady states, we calculate the dependence of the expectation value of the tunneling $\langle K\rangle/L$ on the tunneling amplitude $J$ within the effective model and solve numerically the self-consistency condition. Additionally to the existence of non-trivial solutions, their stability needs to be assured. 

A stability condition can be derived by probing the dynamics of the cavity field and only considering its linear fluctuations above the desired solution similar to the approach in Refs.~\cite{Tian2016,RitschEsslinger2013}. Using Eq.~(\ref{eq:a_time}), the equations of motion of the coordinate and momentum quadratures of the cavity field, i.e., $x_a=\langle a+a^\dagger\rangle$ and $p_a=-i\langle a-a^\dagger\rangle$ are 
\begin{align}
\label{eq:quadratures}
&\partial_t x_a=-\kappa x_a+\delta_{cp}p_a\\
&\partial_t p_a=-\delta_{cp} x_a-\kappa p_a+2\tilde{\Omega} \langle K \rangle \nonumber
\end{align}
with stationary solutions $x_a^{(s)}=\frac{2\delta_{cp}\tilde{\Omega}\langle K \rangle^{(s)}}{\delta_{cp}^2+\kappa^2}$ and $p_a^{(s)}=\frac{2\kappa\tilde{\Omega}\langle K \rangle^{(s)}}{\delta_{cp}^2+\kappa^2}$. 
We consider linear fluctuations around the stationary solutions, $x_a=x_a^{(s)}+\tilde{x}_a$ and $p_a=p_a^{(s)}+\tilde{p}_a$, and also linearize the average of the tunneling in terms of the fluctuations
$\langle K \rangle\approx \langle K \rangle^{(s)}+\frac{d\langle K\rangle^{(s)}}{dx_a^{(s)}}\tilde{x}_a\;$, where $\langle K \rangle^{(s)}$ is the value of the tunneling corresponding to the stationary solution $x_a^{(s)}$.
From Eqs.~(\ref{eq:quadratures}) we can derive a set of differential equations for the fluctuations
\begin{align}
\label{eq:fluctuations}
&\partial_t \tilde{x}_a=-\kappa \tilde{x}_a+\delta_{cp}\tilde{p}_a\\
&\partial_t \tilde{p}_a=\left(-\delta_{cp}+2\tilde{\Omega}\frac{d\langle K \rangle^{(s)}}{dx_a^{(s)}}\right)\tilde{x}_a-\kappa \tilde{p}_a. \nonumber
\end{align}
The eigenvalues of the Jacobian of this set of differential equations are given by 
\begin{equation}
\lambda_{\pm}=-\kappa\pm\sqrt{\left(-\delta_{cp}^2+2\delta_{cp}\tilde{\Omega}\frac{d\langle K \rangle^{(s)}}{dx_a^{(s)}}\right)}.
\end{equation}

The stable stationary solutions are the ones for which the eigenvalues have a negative real part. 

Thus, the stability condition for the system with $\delta_{cp}>0 $ reads 
\begin{equation}
\left(\frac{d\langle K\rangle^{(s)}/L}{dJ^{(s)}/|U|}\right) < \frac{|U|}{AL}
\end{equation}
where the derivative of the tunneling is evaluated at the stationary solution.

The model (Eq.~\ref{eq:Lindblad}) possesses a $Z_2$ symmetry, since it is invariant under the transformation $a\rightarrow -a, c_{j,\sigma}\rightarrow (-1)^j c_{j,\sigma}$. For the parameters under consideration, the tunneling $K$ has the same sign as $J$, such that there exists only a non-trivial steady state if the cavity-pump detuning $\delta_{cp}$ is positive (blue detuned). Without loss of generality we consider $J \geq 0$ in our calculations. 

We determine the physical properties of the effective model using the density matrix renormalization group (DMRG) algorithm. A high-performance DMRG code for finite systems with open boundary conditions which uses ITensor library \cite{itensor}, enables us to target correlations over long distances.  We focus in our simulations on a filling $n=\frac{N_\uparrow+N_\downarrow}{L}=0.9375$  and zero magnetization $M_z=N_\uparrow-N_\downarrow=0$ where $N_\uparrow$ ($N_\downarrow$) is the number of fermions with spin up (down) and $L$ is the number of sites of the chain. We expect the main findings to be stable with respect to parameter changes. In DMRG simulations we use a chain of $L = 192$ with $N_\uparrow+N_\downarrow=180$ or $L = 384$ with $N_\uparrow+N_\downarrow=360$. The results presented in this work are calculated using a bond dimension of up to $M=5000$ which leads, as we verified, to a good accuracy for the considered cases. 

\section{Self-organized s-wave superconducting state}\label{sec:swave}

%-----------------------------------------------------------------------------------------------
% Fig. 02
%-----------------------------------------------------------------------------------------------
\begin{figure}
\includegraphics[width=0.9\linewidth]{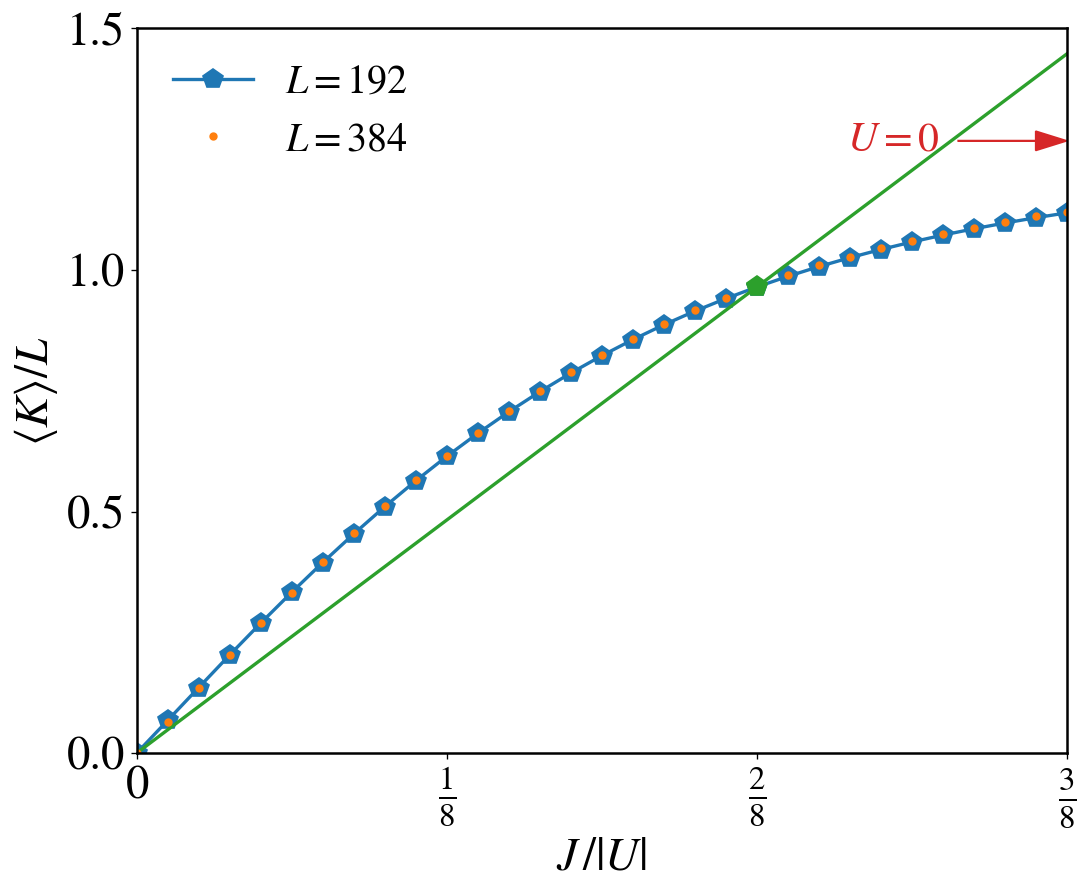}
\caption{\label{fig:Kattra}(color online) The expectation value of the tunneling $\langle K\rangle/L$ for an attractive Hubbard chain $ U<0$ versus the rescaled tunneling amplitude $J/|U|$. Shown are two system size $L=192, L=384$ with filling $n=0.9375$ which lie on top of each other. When the tunneling is very large compared to the interaction the value of the tunneling approaches the non-interacting value ( $U=0$) shown by the red arrow. The green solid line is the linear curve with slope $\left(\frac{AL}{|U|}\right)^{-1}\approx 3.87$. The crossing of the two curves gives the self consistent solution with $J = \frac{2}{8} |U|$.
}
\end{figure}
%----------------------------------------------------------------------------------------------- 
%-----------------------------------------------------------------------------------------------
% Fig. 03
%-----------------------------------------------------------------------------------------------
\begin{figure}
\includegraphics[width=0.9\linewidth]{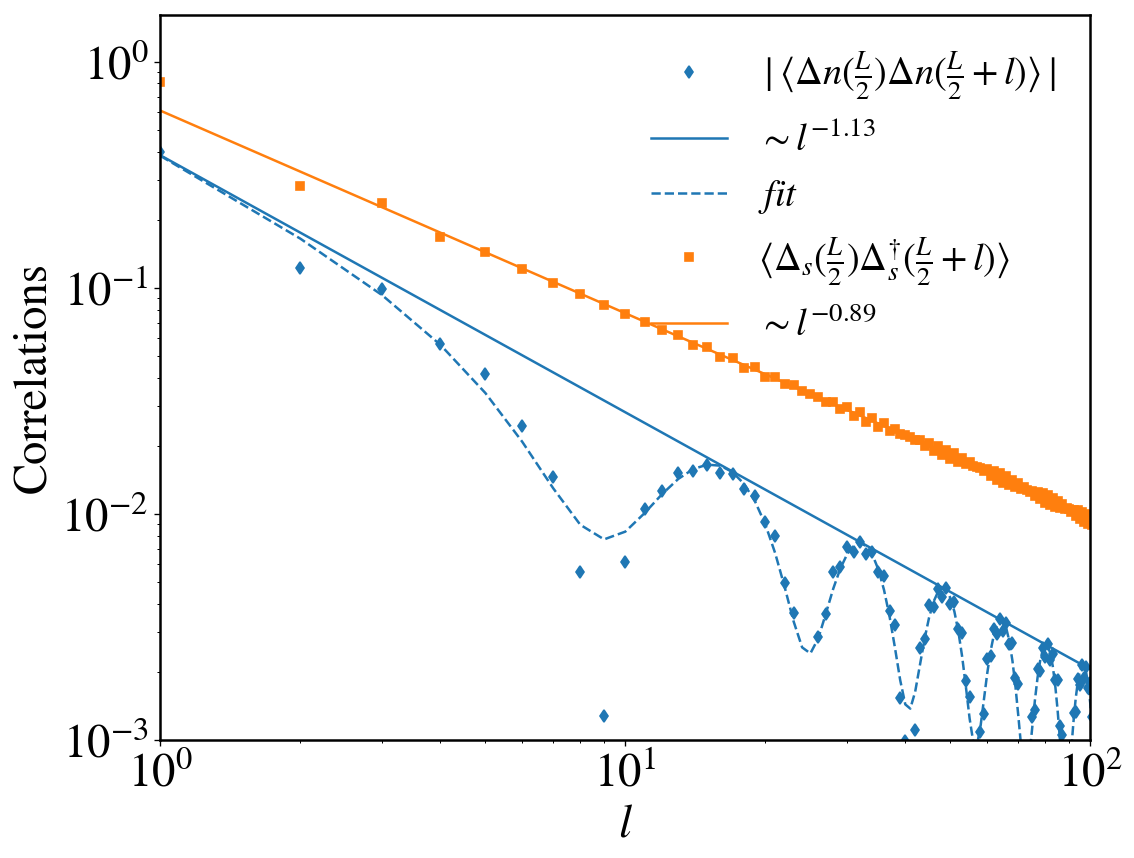}
\caption{\label{fig:swavecorr}(color online) The density-density and s-wave pairing correlations of the attractive Hubbard chain with $ J=\frac{2}{8}|U|$ calculated from DMRG for a system size $L=384$ and $N=360$ particles ($n=0.9375$). Whereas both correlations decay algebraically, the s-wave pairing correlation is the dominant correlation. The pairing correlation has oscillations with period of $2\pi n=4k_F$ on top and is fitted with the function $l^{-\gamma} (a + b \cos(2\pi n(l-l_0 )))$ (dashed line). Fits with algebraic functions  $\sim l^{-\gamma} $ are also shown (solid lines). } 
\end{figure}
%----------------------------------------------------------------------------------------------- 

In order to determine the non-trivial self-consistent solutions, we calculated using DMRG methods the dependence of the expectation value of the tunneling $\langle K \rangle$ on the tunneling amplitude $J$ and solve the self-consistency condition numerically. 
In Fig.~\ref{fig:Kattra} the expectation value of the tunneling $\langle K \rangle/L$ is represented for different tunneling amplitudes rescaled by the interaction. The self-consistency condition can be rewritten as $\langle K\rangle/L = \frac{|U|}{AL} J/|U|$. This form makes it clear that the self-consistency condition has an easy graphical interpretation. 
Using the dependence of the expectation value of the tunneling on the tunneling amplitude shown in Fig.~\ref{fig:Kattra}, the condition can be interpreted graphically by determining the intersection of this curve  $\langle K\rangle/L$ with a linear curve $\frac{|U|}{AL} J/|U|$ where its slope $\left(\frac{AL}{|U|}\right)^{-1}$ depends on the pump strength $A$. Moreover the graphical interpretation of the stability condition is that a solution is stable if at the corresponding intersection the derivative of the curve for the expectation value of the tunneling is less than the slope of the linear curve. For large values of the pump strength there exists a stable non-trivial self-consistent solution of the self-consistency equation as shown for one example in Fig.~\ref{fig:Kattra}. The exact dependence of the expectation value of the tunneling at low values of the tunneling amplitude $J$ determines whether there exists a critical pump strength below which only an empty cavity field is the trivial solution of the self-consistency equation. Our numerical solution suggests that above a critical pump strength the non-trivial solutions emerge. Due to the finite resolution of the numerical data and the slight finite size effects, we cannot pinpoint an exact value of this critical pump strength.  
The non-trivial self-consistent solutions at large pump strength $A$ are stable, since the slope of the expectation value of the tunneling $\aver{K}/L$ versus the tunneling amplitude $J/|U|$ is smaller than the slope $\left(\frac{AL}{|U|}\right)^{-1}$ and fulfills thus the stability condition. 
In the non-trivial self-consistent solution, the pump photons scatter via the cavity-assisted tunneling into the cavity and lead to a finite cavity occupation. In this steady state the fermions organize into an s-wave superconducting phase which is formed by the pairs of spin-up and spin-down fermions. The properties of this s-wave superconducting phase have been investigated previously for the attractive Hubbard model (see for example \cite{Giamarchibook, EsslerKorepin2005}). In one-dimension, this superconducting phase is characterized by dominating s-wave pairing correlations defined by 
$\langle \Delta_s(j) \Delta_s^\dagger(j+l) \rangle$ with the onsite pairs  $\Delta_s(j)=c_{j,\uparrow}c_{j,\downarrow}-c_{j,\downarrow}c_{j,\uparrow}$ which decay algebraically with distance. This is shown in Fig.~\ref{fig:swavecorr} for an attractive interaction with $ J=\frac{2}{8}|U|$. Compared to other correlations in the system the decay of the s-wave pairing correlation is the slowest decay and it oscillates with the period of $2\pi n= 4 k_F$. Due to the gap in the spin sector of the excitations, the spin correlations decay exponentially fast with distance (not shown) and the faster algebraic decay of the density-density correlations ($\Delta n(j) =n(j)-\langle n(j)\rangle$) is shown in Fig.~\ref{fig:swavecorr}. We note that in the weakly interacting regime the bosonization method shows a relation between the exponent of the algebraic decay of these correlations; the density-density correlation decays as $l^{-K_\rho}$ in which $K_\rho$ is known as the Luttinger liquid parameter while the s-wave pairing correlation decays as $l^{-1/K_\rho}$. For the parameters used in Fig.~\ref{fig:swavecorr} the Luttinger liquid parameter extracted from the density-density correlations $K_\rho = 1.13$ is very close to the Luttinger liquid parameter extracted from the s-wave pairing correlations $K_\rho = \frac{1}{0.89}=1.124$. One would expect that in the thermodynamic limit ($L\to \infty$) these two values become equal. The cavity-induced s-wave superconducting phase is stable by the cavity dissipation and leaking of photons from the cavity signals the emergence of the super-radiant phase and equivalently the superconducting phase.

\section{Setup and theoretical model for the ladder structure}\label{sec:setupq1d}
%-----------------------------------------------------------------------------------------------
% Fig. 04 setup ladder
%-----------------------------------------------------------------------------------------------
\begin{figure}
  \includegraphics[width=0.7\linewidth]{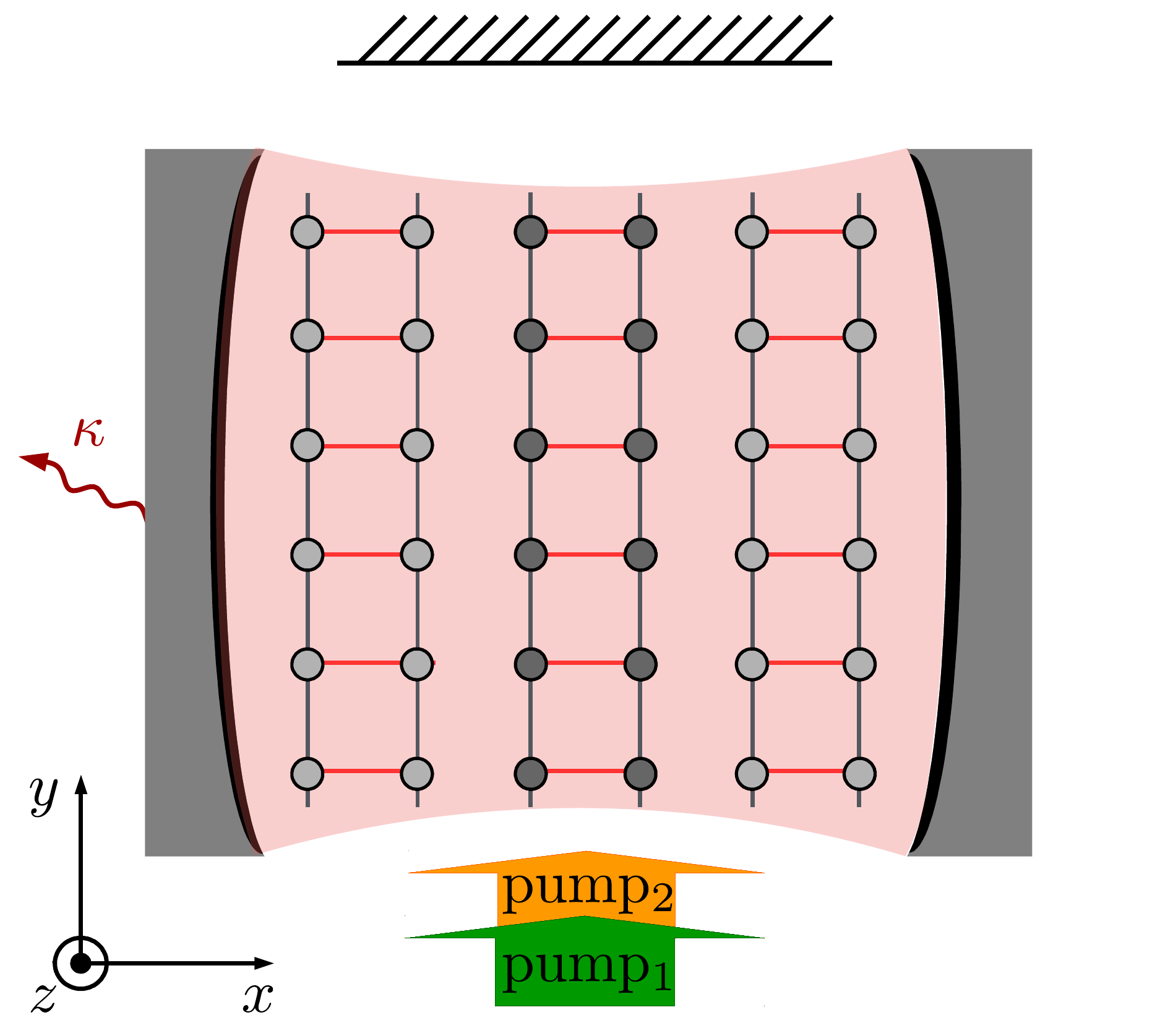}
\caption{\label{fig:setup2}(color online) A balanced mixture of fermionic atoms in two internal states is loaded into a structure of decoupled ladders formed by optical lattices. Tunneling along the rung of the ladder is strongly suppressed by a potential offset $\Delta$ between neighboring lattice sites and restored by a cavity-assisted Raman process.} 
\end{figure}
%----------------------------------------------------------------------------------------------- 

%-----------------------------------------------------------------------------------------------
% Fig. 05
%-----------------------------------------------------------------------------------------------
\begin{figure}
\includegraphics[width=0.9\linewidth]{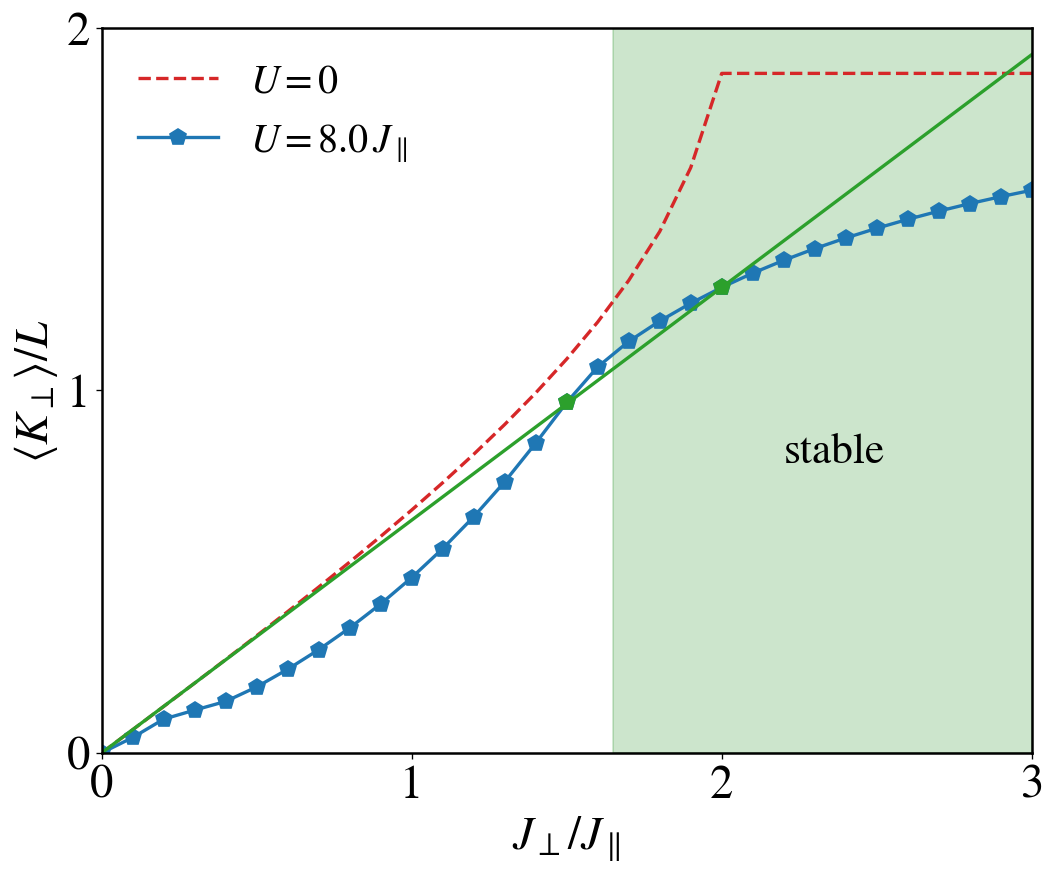}\\
\caption{\label{fig:Kperp}(color online) Graphical interpretation of the self-consistency condition. The dependence of the rung tunneling (blue circles) on different ratios of the tunneling amplitudes $J_\perp/J_\parallel$ is shown for the repulsive Hubbard ladder with $ U =8J_\parallel$ in a system of $L=192$ rungs and $N=360$ particles ($n=0.9375$) and for the non-interacting Hubbard ladder (red dashed line). The intersections with the linear function (green solid line) $\left(\frac{AL}{J_\parallel}\right)^{-1}\approx 0.64$ give the solutions of the self-consistency condition. 
For $U=8 J_\parallel$, the intersection at $J_\perp= 2 J_\parallel$ indicates a stable non-trivial self consistent solution. The intersection around $J_\perp= 1.5 J_\parallel$ is not stable. The green shaded region marks the regime of possible stable solutions.}
\end{figure}
%----------------------------------------------------------------------------------------------- 
%-----------------------------------------------------------------------------------------------
% Fig. 06
%-----------------------------------------------------------------------------------------------
\begin{figure}
\includegraphics[width=0.9\linewidth]{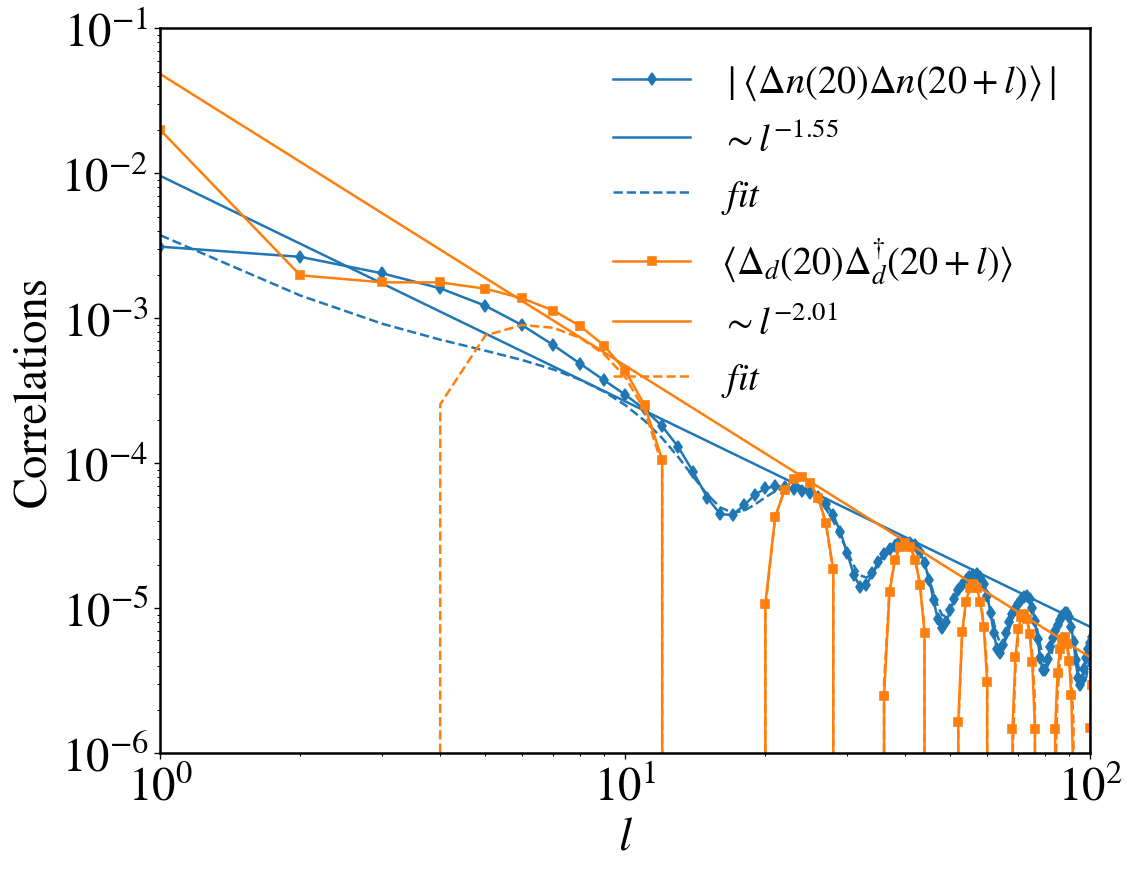}
\caption{\label{fig:dwavecorr}(color online) The density-density correlations (orange squares) and the unconventional pair correlations (blue circles) versus distance $l$ computed starting from site $j=20$ in a Hubbard ladder with $ U =8J_\parallel, J_\perp=2 J_\parallel $ for systems of $L=192$ rungs and $N=352$ particles ($n=0.9375$). Both correlations decay algebraically with distance. The curves are fitted with the function $l^{-\beta} (a + b \cos(2\pi n(l-l_0 )))$ with the period of $2\pi n = 4k_F$ (dashed lines). The fits with only the algebraic function are also shown as guide to the eye (solid lines). The density-density correlations decay slower than the superconducting correlations and, thus, are dominating which is the signature of the charge density wave CDW$^{4k_F}$.}
\end{figure}
%----------------------------------------------------------------------------------------------- 

We extend the model to an array of ladders in which the tunneling along the legs of the ladders is the standard tunneling and only the tunneling on the rungs of the ladders is induced by the Raman process (see Fig.~\ref{fig:setup2}). In order to reach this situation, decoupled ladders with an energy offset $\Delta$ between their legs are generated by three dimensional optical (super)-lattices (see for example Ref.~\cite{SheikhanKollath2016} for more details). No energy offset along the legs of the ladders is employed. Thus, the tunneling along the legs is the standard tunneling and the tunneling along the rungs of the ladders is initially suppressed and reenforced by the cavity-assisted tunneling introduced in the previous section. Performing the same approximations as before, the Hamiltonian describing the ladder structures reads
\begin{eqnarray}\label{eq:Ham_i_ladder}
&&H = H_c + H_\parallel + H_{ac} + H_{\text{int}}, \nonumber\\
&&H_c=\hbar \delta_{cp} a^\dagger a, \nonumber\\
&&H_\parallel=-J_\parallel \sum_{j,\sigma,m=0,1} \left( c^\dagger_{m,j,\sigma} c_{m,j+1,\sigma}+\mathrm{H.c.}\right), \nonumber\\
&&H_{ac}= - \hbar\tilde{\Omega} (a^\dagger+a) \sum_{j,\sigma} \left (c^\dagger_{0,j,\sigma} c_{1,j,\sigma}+\mathrm{H.c.} \right ),\nonumber\\
&&H_{\text{int}}= U \sum_{j,m=0,1} n_{m,j,\uparrow} n_{m,j,\downarrow}.
\end{eqnarray}
Here, $c_{m, j,\sigma}$ $(c_{m,j,\sigma}^{\dag})$ annihilates (creates) an atom with spin $\sigma=\uparrow,\downarrow$ on rung $j$ and leg $m$ of the ladder. For notational simplicity we introduce the tunneling operator along the rungs of the ladder as $K_\perp = \sum_{j,\sigma}\left ( c^\dagger_{0,j,\sigma} c_{1,j,\sigma} +\mathrm{H.c.}\right )$. The dynamics of the system is again described by a Lindblad master equation due to the presence of losses from the cavity. Using the adiabatic elimination of the cavity field, one can derive an effective Hamiltonian for the fermions with a self-consistency condition given by,

\begin{eqnarray}\label{eq:Ham_F_ladder}
&&H_{F} = H_\parallel +H_\perp + H_{\text{int}}, \nonumber\\
&&H_\perp= - J_\perp K_\perp, \nonumber\\
&&J_\perp=A \langle K_\perp \rangle.
\end{eqnarray}
The pump strength $A= \frac{2\hbar\tilde{\Omega}^2\delta_{cp}}{\delta_{cp}^2 +\kappa^2}$ is defined as before. The effective Hamiltonian is the Hubbard Hamiltonian on a ladder which shows the power-law decay of the correlation functions and has been studied before, see eg. \cite{Giamarchibook,OrignacGiamarchi1997} for analytical and \cite{NoackScalapinto1994,NoackScalapino1995,NoackScalapino1996,DolfiTroyer2015} for numerical studies.
The stability condition for the solutions is $\left(\frac{d\langle K_\perp\rangle^{(s)}/L}{dJ_\perp^{(s)}/J_\parallel}\right)< (\frac{AL}{J_\parallel})^{-1}$.

We simulate the effective model employing the DMRG method for a ladder of size $2\times L$ with $L=192$, $N=360$ particles and $M_z=0$. The maximum bond dimension we consider is $M=5000$ and we checked the convergence of our results in this parameter.

\section{Self-organization of the $4k_F$ charge density wave}
\label{sec:cdw}

%%%%%%%%%%self-consistent solution
As in the previous situation, a graphical interpretation of the self-consistency condition $\langle K_\perp\rangle/L = \frac{J_\parallel}{AL} J_\perp/J_\parallel$ exists. 
Graphically, this means that in order to find the solutions of the self-consistency equation one needs to find the intersections of $\langle K_\perp \rangle/L$ and a linear line with the slope $(\frac{AL}{J_\parallel})^{-1}$. In Fig.~\ref{fig:Kperp} the expectation value of the rung tunneling $\langle K_\perp \rangle/L$ is plotted versus the tunneling amplitude $J_\perp/J_\parallel$. A monotonic rise with a turning in the curvature is found. 
Due to this complex dependence of the expectation value of the rung tunneling on the tunneling amplitude, different situations can occur depending on the value of the pump strength $A$: (i) there might be no intersection, i.e.~no non-trivial solution exists, (ii) one intersection, (iii) more than one intersection for a certain pump strength (shown). 
Our numerical solution suggests a critical pump strength below which no non-trivial stable solution of the self-consistency condition exists, i.e.~in this steady state the cavity is not occupied and no tunneling between the different legs of the ladders occurs. Above the critical value in an interval of the pump strength $A_{cr,1}<A<A_{cr,2}$ our numerical solution gives at least two solutions of the self-consistency condition. However, in the case of two non-trivial solutions only the solution with the larger value of the tunneling amplitude is stable according to the stability condition. For strong values of the pump strength, $A>A_{cr,2}$, our results show only one stable non-trivial solution. As shown in Fig.~\ref{fig:Kperp} the solutions with $J_\perp/J_\parallel > 1.65$ are stable.

For each of the stable non-trivial solutions the cavity field becomes dynamically occupied and the two legs of the ladder are coupled by the cavity-induced tunneling. In order to obtain information about the properties of the fermions in this steady state, we need to consider the properties of the effective Hamiltonian. At repulsive interaction in the Hubbard model on a ladder, a crossover between two interesting phases occurs. At intermediate rung tunneling, an unconventional superconductor emerges and at small and large rung tunneling a $4k_F$ charge density wave (CDW$^{4k_F}$) forms \cite{Giamarchibook,OrignacGiamarchi1997}. $k_F$ is the Fermi wave-vector which is set by the filling as $2\pi n = 4 k_F $.

Both the singlet-pair correlations, $\langle \Delta_d(j) \Delta_d^\dagger(j+l)\rangle $ with the singlet on a rung $\Delta_d(j)=c_{0,j,\uparrow}c_{1,j,\downarrow}-c_{0,j,\downarrow}c_{1,j,\uparrow}$, and the density-density correlations, $\langle \Delta n(j) \Delta n(j+l) \rangle$, decay algebraically. The slowest and, thus, the dominating decay is characterizing the properties of the state. The algebraic decay of the correlations in the Hubbard ladder has been studied previously using DMRG \cite{NoackScalapinto1994,NoackScalapino1995,NoackScalapino1996,DolfiTroyer2015}.

For the considered interaction strength of $U=8 J_\parallel$ we determine the crossover between d-wave superconductor and $4k_F$-charge-density wave by comparing the algebraic decay of the singlet-pair correlations, i.e.~$\langle \Delta_d(j) \Delta_d^\dagger(j+l)\rangle$,  and density-density correlations for different rung tunneling. As an example, we show in Fig.~\ref{fig:dwavecorr} the density correlation and singlet-pair correlations for the solution with $J_\perp/J_\parallel = 2$. The period of $4k_F = 2\pi n $ is clearly seen in both correlations. Whereas at small distances the singlet-pair correlation has a larger amplitude than the density-density correlation, at long distances the $4k_F$-density-density correlation shows the slower decay and is dominating. The bosonization predicts that the exponent of the d-wave pairing correlation is equal to the inverse of the exponent of the density-density correlation which is not yet the case for the results shown in Fig.~\ref{fig:dwavecorr}. One would expect that this relation is fulfilled in the thermodynamic limit and here we can discuss the slower decay of the d-wave pairing correlations compared to other correlations up to the system size considered in the simulations ($L=192$).     

The crossover between the phases approximately occurs at $\left(J_\perp/J_\parallel\right)_{cr} \approx 1.6 $ with a dominating CDW$^{4k_F}$ phase for larger values of $J_\perp/J_\parallel$ which corresponds to the stable non-trivial solutions of the self-consistency condition (see Fig.~\ref{fig:Kperp}). This means that in the self-organized setup at the considered interaction strength only the CDW$^{4k_F}$ phase is stabilized. Let us note that we find a similar behavior for different interaction strength, different fillings and different anisotropies of tunneling along the legs of the ladder. Thus, it seems that the unconventional superconducting state cannot be stabilized easily by the coupling to the cavity.

\section{Conclusion}
To summarize, the coupling mechanism we introduced between atoms and the cavity field can lead to a self organization into an s-wave superconducting phase for a chain of attractively interacting atoms and into a $4k_F$-charge density wave on a ladder geometry with repulsive interaction. These steady states are stable and protected against dissipative fluctuations of the system and can be realized in the experiment by the super-radiant phase in which the cavity field is occupied and photons leak from the cavity. We could not stabilize in the ladder geometry the unconventional superconducting phase for the considered parameters even changing the interaction, particle filling and anisotropy of the tunneling along the legs of the ladder. 

\section{Acknowledgements}
We would like to thank T.~Giamarchi, M.~K\"ohl and E.~Orignac for fruitful discussions. 
We acknowledge support from research council of Shahid Beheshti University, G.C. (A.S.), the DFG (TR 185 project B3, FOR1807, SFB 1238 project C05, and Einzelantrag) and the ERC (Grant Number 648166) (C.K.).

%==============================================================================================

\end{document}